\documentclass[iop]{emulateapj}
\usepackage{apjfonts,rotating,eqnarray}
\usepackage{xcolor}
\usepackage{amsmath}

 \def\chandra{{\itshape Chandra\/}}

 \def\hst{{\itshape HST\/}}

 \def\spitzer{{\itshape Spitzer\/}}

 \def\xray{\hbox{X-ray}}

 \def\etal{{et\,al.}}

 \def\ltsima{$\; \buildrel < \over \sim \;$}
 \def\simlt{\lower.5ex\hbox{\ltsima}}
 \def\gtsima{$\; \buildrel > \over \sim \;$}
 \def\simgt{\lower.5ex\hbox{\gtsima}}
 \def\kms{\ifmmode{~{\rm km~s^{-1}}}\else{~km s$^{-1}$}\fi}
 \def\lsim{\lower0.3em\hbox{$\,\buildrel <\over\sim\,$}}
 \def\gsim{\lower0.3em\hbox{$\,\buildrel >\over\sim\,$}}
 
 \def\lksol{$L_{K,\odot}$}
 \def\msol{$M_\odot$}
 \def\h2{H$_2$}
 
 \def\lum{erg~s$^{-1}$}

 \def\apj{ApJ}
 \def\apjl{ApJL}
 \def\apjs{ApJS}
 
 \def\mnras{MNRAS}
 \def\araa{ARA\&A}
 
 \def\nat{Nature}

 \def\PeacockEtAl{P14}

 \def\mstarlow{$0.679$}
 \def\mstarhigh{$3.10$}
 \def\MLlow{$1.48$}
 \def\MLhigh{$4.12$}

 \def\firstalphaone{$2.07$}
 \def\firstalphatwo{$2.20$}
 \def\firstgamma{$1.30$}

 \def\bestalphaone{$3.84$}
 \def\bestalphatwo{$2.14$}
 \def\bestgamma{$1.30$}

 \def\aonebefore{$\alpha_1 =\firstalphaone^{+1.9}_{-1.1}$}
 \def\aoneafter{$\alpha_1 = \bestalphaone^{+0.09}_{-0.48}$}

 \def\atwobefore{$\alpha_2 = \firstalphatwo^{+0.17}_{-0.24}$}
 \def\atwoafter{$\alpha_2 = \bestalphatwo^{+0.20}_{-0.85}$}

 \def\scbefore{$\gamma_{\textup{X}} = \firstgamma^{+0.55}_{-0.45}$ ($10^{10}$~\lksol)$^{-1}$}
 \def\scafter{$\gamma_{\textup{X}} = \bestgamma^{+0.65}_{-0.35}$ ($10^{10}$~\lksol)$^{-1}$}

\begin{document}

\shortauthors{COULTER ET AL.}
\shorttitle{Testing the Universality of the Stellar IMF with \chandra~and \hst}

%
\title{Testing the Universality of the Stellar IMF with {\itshape Chandra} and \hst}
%

\author{
D.~A.~Coulter,\altaffilmark{1,2}
B.~D.~Lehmer,\altaffilmark{3}
R.~T.~Eufrasio,\altaffilmark{2,3}
A.~Kundu,\altaffilmark{4}
T.~Maccarone,\altaffilmark{5}
M.~Peacock,\altaffilmark{6}
A.~E.~Hornschemeier,\altaffilmark{2}
A.~Basu-Zych,\altaffilmark{2}
A.~H.~Gonzalez,\altaffilmark{7}
C.~Maraston,\altaffilmark{8}
\& S.~E.~Zepf\altaffilmark{5}
}

\altaffiltext{1}{University of California Santa Cruz, Santa Cruz, CA
95064, USA}
\altaffiltext{2}{NASA Goddard Space Flight Center, Code 662, Greenbelt, MD
20771, USA} 
\altaffiltext{3}{Department of Physics, University of Arkansas, 226 Physics
Building, 835 West Dickson Street, Fayetteville, AR 72701, USA}
\altaffiltext{4}{Eureka Scientific, Oakland, CA 94602, USA} 
\altaffiltext{5}{Texas Tech University, Lubbock, TX 79409, USA} 
\altaffiltext{6}{Michigan State University, East Lansing, MI 48824, USA} 
\altaffiltext{7}{University of Florida, Gainesville, FL 32611, USA} 
\altaffiltext{8}{University of Portsmouth, Portsmouth, United Kingdom} 
%

%
\begin{abstract}
%

The stellar initial mass function (IMF), which is often assumed to be universal
across unresolved stellar populations, has recently been suggested to be ``bottom-heavy'' 
for massive ellipticals. In these galaxies, the prevalence of gravity-sensitive 
absorption lines (e.g. Na~I and Ca~II) in their near-IR spectra implies an excess of 
low-mass ($m\simlt0.5$~\msol) stars over that expected from a canonical IMF observed in 
low-mass ellipticals. A direct extrapolation of such a bottom-heavy IMF to high stellar 
masses ($m\simgt8$~\msol) would lead to a corresponding deficit of neutron stars and black holes, 
and therefore of low-mass \xray\ binaries (LMXBs), per unit near-IR luminosity in these 
galaxies. Peacock et al.~(2014) searched for evidence of this trend and found that the 
observed number of LMXBs per unit $K$-band luminosity ($N/L_K$) was nearly constant. We 
extend this work using new and archival {\itshape Chandra X-ray Observatory} 
(\chandra\/) and {\itshape Hubble Space Telescope} (\hst\/) observations of seven low-mass 
ellipticals where $N/L_K$ is expected to be the largest and compare these data with a 
variety of IMF models to test which are consistent with the observed $N/L_K$. 
We reproduce the result of Peacock et al.~(2014), strengthening the constraint that 
the slope of the IMF at $m\simgt8$~\msol\ must be consistent with a Kroupa-like IMF. We construct  
an IMF model that is a linear combination of a Milky Way-like IMF and a broken power-law IMF, 
with a steep slope ($\alpha_1=$ $3.84$) for stars < 0.5~\msol~(as suggested
by near-IR indices), and that flattens out ($\alpha_2=$~$2.14$) for stars > 0.5~\msol, and discuss 
its wider ramifications and limitations.

\end{abstract}
%


%
\section{Introduction}
%

\noindent Understanding the stellar initial mass function (IMF) has major
implications for a variety of astrophysical problems.  The IMF plays a central
role in converting the observed properties of galaxies (e.g., luminosity and
color) into physically meaningful quantities like stellar mass and
star-formation rate.  An assumed universality of the IMF has contributed to our
current observational-based paradigm of how galaxies formed and evolved
throughout the history of the Universe, what fraction of the Universe's mass is
tied up in stellar baryons, and the number of compact objects in the Universe. A
clear understanding of the form and universality of the IMF in external
galaxies is therefore a central goal in modern astrophysics (see, e.g., Bastian
\etal\ 2010 for a review).

In recent years, there has been mounting evidence indicating that the stellar
IMFs for elliptical galaxies vary with galaxy mass. For example, Cappellari
\etal\ (2012) used detailed dynamical models and two-dimensional stellar
kinematic maps of 260 early-type galaxies from the ATLAS$^{\textup{3D}}$
project to show that the mass-to-light ratios ($M/L$) for elliptical galaxies
increase with increasing stellar velocity dispersion, $\sigma$, consistent with
a scenario where the IMF changes with galaxy mass.  Such a finding has been
independently noted in galaxy lensing measurements of $M/L$ for a variety of
$\sigma$ (e.g., Auger \etal\ 2010; Treu \etal\ 2010). Specifically, these
findings indicate that relatively low-mass, early-type galaxies (\hbox{$\sigma \simlt
100$~km~s$^{-1}$}) have $M/L$ values consistent with standard Milky Way-like IMFs (Kroupa 2001; 
Chabrier 2003). However, relatively massive ellipticals (\hbox{$\sigma \approx
300$~km~s$^{-1}$}) have mass-to-light ratios ($M/L$) that are larger
than those predicted from standard IMFs, and can be consistent with either ``bottom-heavy'' IMFs 
(\hbox{$\alpha \approx 2.8$}, yielding more low-mass stars with higher $M/L$) or ``top-heavy'' 
IMFs (\hbox{$\alpha  \approx 1.5$}, yielding more remnants with lower luminosities; Cappellari et al. 
2012).


\begin{table*}
\begin{center}
\footnotesize
\caption{Properties of Low-Mass Elliptical Galaxy Sample}
\begin{tabular}{lccccccccccccccc}
\hline\hline
& $D$ & $\sigma$ & $a$ & $b$ & $\log L_K$ & $N_{\textup{H}}$ & \multicolumn{2}{c}{\textit{HST} ACS Data} & $t_{\rm exp}$ & $N_{\rm LMXB}$ & $N_{\rm X, GC}$ & $N_{\rm X, bkg}$ \\
 \multicolumn{1}{c}{Source Name} & (Mpc) & (km~s$^{-1}$) & \multicolumn{2}{c}{(arcmin)} & ($\log$ \lksol) & ($10^{20}$ cm$^{-2}$) & (Blue Filter) & (Red Filter) & (ks) & (field) & (GCs) & (Background) \\
 \multicolumn{1}{c}{(1)} & (2) & (3) & (4) & (5) & (6) & (7) & (8) & (9) & (10) & (11) & (12) & (13) \\
\hline
        NGC 4339 \ldots\ldots\ldots\ldots 	&  16.0 & 100.0 &   1.3 &   1.1 &  10.3$^{\dag}$ & 1.62 & F606W & ---  & 33.6 &  2$^{\dag}$  &  0$^{\dag}$ & 1$^{\dag}$ \\
        NGC 4387 \dotfill	&  17.9 &  97.0 &   0.9 &   0.6 &  10.2 & 2.73 & F475W & F850LP &  38.7 &  1  &  0 & 1 \\
        NGC 4458\dotfill 	&  16.4 &  85.0 &   0.9 &   0.7 &  10.0 & 2.63 & F475W & F850LP &  34.5 &  2  &  0 & 1 \\
        NGC 4550\dotfill 	&  15.5 & 110.0 &   1.3 &   0.5 &  10.2 & 2.60 & F475W & F850LP &  25.8 &  6  &  1 & 1 \\
        NGC 4551\dotfill 	&  16.1 &  95.0 &   1.1 &   0.7 &  10.2 & 2.59  & F475W & F850LP &  26.6 &  0  &  1 & 0 \\
        NGC 7457$^*$\dotfill &  12.9 &  78.0 &   2.6 &   1.4 &  10.3 & 5.49 & F475W & F850LP &  37.7 &  1  &  0 & 2 \\
\hline
Total Sample\dotfill & \multicolumn{9}{c}{ } & {\bf 12} &  2 & 6 \\
\hline
\end{tabular}
\end{center}
{\scriptsize NOTE.---{\itshape Col.(1)}: Target galaxy name.  {\itshape Col.(2)}: Distance as
given by Cappellari \etal\ (2013). {\itshape Col.(3)}: Velocity dispersion from either
Halliday \etal\ (2001) or Cappellari \etal\ (2006). {\itshape Col.(4)} and {\itshape (5)}: 2MASS
based $K$-band galaxy major and minor axes from Jarrett \etal\ (2003). {\itshape Col.(6)}:
Logarithm of the $K$-band luminosity.  {\itshape Col.(7)}: Neutral hydrogen Galactic
column density. {\itshape Col.(8)} and {\itshape (9)} The available $HST$ imaging (via either ACS or
WFPC2 imaging) for ``blue'' and ``red'' filters, respectively, which were used
to identify and characterize optical counterparts to \xray\ sources.  {\itshape Col.(10)}:
Total \chandra\ exposure time.  All galaxies were imaged using ACIS-S and had
coverage over the entire galactic extents as defined in {\itshape Col.~(4)} and {\itshape (5)}.
{\itshape Col.(11)--(13)}: The number of field LMXBs ({\itshape Col.(11)}), GC LMXBs ({\itshape Col.(12)}), and
background or central AGN candidates ({\itshape Col.(13)}) within the galactic extent of
each galaxy (as defined in {\itshape Col.(4)} and {\itshape (5)}) that had \hbox{0.5--7~keV} fluxes
exceeding that of a $L_{\rm X} = 10^{38}$~\lum\ source at the distance provided
in {\itshape Col.(2)}.  \xray\ sources were classified using \hst\ data and the procedure
outlined in \S2.2 of \PeacockEtAl.\\
\noindent $^*$Observations of NGC 7457 were represented in both the sample by
\PeacockEtAl~and this study. However, due to the availability of new
\hst\ data for NGC 7457 in this study, results based on our analysis were used
throughout this paper.\\
\noindent \dag~For NGC 4339, we considered 74\% of the reported $K$-band
luminosity that corresponded to the region of the galaxy covered by the \hst\
WFPC2 footprint detailed in \S2. For {\itshape Col.(11)--(13)}, we only considered X-ray sources 
detected in this covered region.
}
\end{table*}

Van Dokkum \& Conroy (2010, 2011, 2012) and Conroy \& van~Dokkum (2012) showed
that the spectra of massive ellipticals have strong Na~I and Ca~II absorption features 
that are indicative of a large population of very low-mass stars ($\simlt$0.3~\msol; 
Saglia et al. 2002), favoring the bottom-heavy IMF interpretation for these galaxies. 
These features had been known for quite some time (see, e.g. Hardy \& Couture 1988; Cenarro 
\etal\ 2003), but higher-quality data and spectral stellar synthesis modeling in 
recent years has given more confidence to the interpretation that the stellar IMFs 
in these galaxies are likely to be bottom-heavy and consistent with $\alpha \approx
2.8$. 

Despite the above evidence for a bottom-heavy IMF in massive
ellipticals, some inconsistencies remain. For example, Smith and Lucey (2013), 
found that a bottom-heavy IMF is incompatible with the stellar mass obtained via 
strong lensing, for a ($\sigma \approx 300$~km~s$^{-1}$) galaxy. Smith (2014) has also 
shown, intriguingly, that there is no correlation between results based on near-IR 
indices and on dynamics, on a galaxy-by-galaxy basis. Additionally, Weidner \etal\ (2013)
showed that a time-invariant and bottom-heavy IMF was incompatible with the
observed chemical enrichment of massive ellipticals, and underpredicted the
number of stellar remnants (in the form of LMXBs) observed in globular clusters
(GCs). To resolve this, they proposed a time-dependent IMF model that would
evolve from a top-heavy to bottom-heavy form. In this scenario, the
star-formation histories of massive ellipticals would have included early
phases of intense starburst activity that produced many massive stars (and a top-heavy IMF), which
chemically enriched the galaxies and created turbulence in their interstellar
mediums.  Subsequent star formation within these environments would be more
greatly fragmented and lead to the formation of preferentially smaller stars
and bottom-heavy IMFs.  

Other authors have taken somewhat different approaches to resolving the above incompatibility
issue.  Ferreras \etal\ (2015) focused on the functional form of the
IMF and tested a model where the low-mass and high-mass slopes of a broken power law IMF were
independently varied. They found that regardless of the IMF slope parameters, a
time-independent IMF (independent of galaxy mass) could not simultaneously
reproduce the observed chemical enrichment and gravity sensitive spectral
features in massive ellipticals.  Finally, Mart\'in-Navarro \etal\ (2014)
explored a radial and time-dependent form of the IMF.  They argued that in
massive ellipticals ($\sigma \approx 300$~km~s$^{-1}$), the radial trends in
chemical enrichment and gravity sensitive spectral features could be well
described by a time-dependent IMF, like that proposed by Weidner \etal\ (2013),
in the central regions of these galaxies. In contrast, a Milky Way-like IMF was
proposed in the galactic outskirts, based on the lack of these spectral
signatures relative to the galactic interior. The IMF was then claimed to be a
local phenomenon, reflecting a two-stage formation history for these galaxies.
Other evidence for radial trends in spectral features follows similar rationales
(La~Barbera \etal\ 2016a, 2016b).


The studies above have helped to constrain variations in the low-mass end of
the elliptical-galaxy IMF ($\simlt$0.5~\msol) with galaxy mass but do not
place strong constraints on variations of the high-mass end of the IMF. To trace 
the high-mass end of the IMF requires studying the most massive stars in a galaxy, 
however these stars evolve quickly into black holes (BHs) and neutron stars (NSs) making them 
difficult to observe. Fortunately these stellar remnants are found in binary star 
systems, where a lower-mass and longer-lived companion star eventually becomes a ``donor'' star, 
losing mass to the compact object. The system itself radiates nearly all of its 
energy in the form of X-rays, making these previously undetectable high-mass remnants 
available to direct observation. Therefore, an efficient and effective way to constrain variations in the 
high-mass end of the IMF ($\simgt$8~\msol) with galaxy mass is to test for
corresponding variations in the prevalence of these low-mass \xray\ binary (LMXB)
systems in elliptical galaxies. With this in mind, Peacock \etal\ (2014; hereafter, \PeacockEtAl)
calculated (using Maraston (2005) population synthesis calculations) that a bottom-heavy 
single power-law IMF, with $\alpha \approx 2.8$, would yield a factor of \hbox{$\approx$3} 
times fewer LMXBs per unit $K$-bandluminosity ($N/L_K$) over a standard Kroupa~(2001) IMF, implying a dramatic
decline in $N/L_K$ with increasing $\sigma$.  Using archival {\itshape Chandra X-ray
Observatory} (\chandra\/) and {\itshape Hubble Space Telescope} (\hst\/) data of eight 
nearby elliptical galaxes, \PeacockEtAl~found instead that
$N/L_K$ was constant with elliptical galaxy luminosity and velocity dispersion;
however, their test included only one low-mass elliptical galaxy at $\sigma <
150$~km~s$^{-1}$, where $N/L_K$ was expected to be largest.

In this paper, we augment the \PeacockEtAl~study of eight elliptical
galaxies by adding \chandra\ constraints for five new low-mass ellipticals
($\sigma =$~78--110~\kms) that are predicted to have standard IMFs that differ
from those of the massive ellipticals (see, e.g., Cappellari
\etal\ 2012). With this expanded sample of 13 galaxies, we are able to place
robust constraints on how $N/L_K$ varies across the full range of velocity
dispersion ($\sigma =$~80--300~km~s$^{-1}$) where significant variations in
$N/L_K$ are plausibly expected.

%
\section{Sample and Data Reduction}
%

Our low-mass elliptical galaxy sample was derived from the samples of Halliday \etal\
(2001) and Cappellari \etal\ (2006), which contain velocity dispersion data for
several nearby ellipticals.  We limited our sample to galaxies with $D <
20$~Mpc, so that we could easily resolve LMXB populations, and $\sigma \simlt
110$~km~s$^{-1}$, to focus on the galaxies that are likely to have standard
IMFs that differ from the already well-studied high-mass ellipticals.  In order
to guard against potential contamination from young high-mass \xray\ binaries (HMXBs), we further
restricted our sample to galaxies that had negligible star-formation rate
signatures (SFR < 0.001 \msol\ yr$^{-1}$), as measured by 24$\mu$m \spitzer\
data (Temi \etal\ 2009).  Our final sample of six low-mass elliptical galaxies
is summarized in Table~1.

We conducted new \chandra\ observations for four of the six galaxies to reach
0.5--7~keV point-source detection limits of $L_{\textup{X}} \approx 10^{38}$
erg~s$^{-1}$ (chosen for consistency with the sample from \PeacockEtAl), after
combining with archival \chandra\ data.  All observations were conducted using
ACIS-S, which covers the entire $K$-band-defined areal footprints of all of the
galaxies in our sample (see Table~1 and Jarrett \etal\ 2003). The only
exception was NGC 4339, which only had partial $HST$ coverage. For this galaxy,
we utilized only the background-subtracted $K$-band flux within the
overlapping $HST$ footprint, and only LMXBs detected in this region were
included in our analyses. This resulted in our using only 74\% of the $K$-band
luminosity reported in Col.(6) when computing $N/L_K$ for this galaxy.

Data reduction for our sample of galaxies closely follows the procedure
outlined in \S\S2.1 and 2.2 of Lehmer \etal\ (2013), with our reduction being
performed with \texttt{CIAO v.~4.7} and \texttt{CALDB v.~4.6.7}.  We
reprocessed events lists from level~1 to level~2 using the script {\ttfamily
chandra\_repro}, which identifies and removes events from bad pixels and
columns, and filters events lists to include only good time intervals without
significant flares and non-cosmic ray events corresponding to the standard ASCA
grade set (grades 0, 2, 3, 4, 6).  For galaxies with more than one observation,
we combined events lists using the script {\ttfamily merge\_obs}.  We
constructed images in three \xray\ bands: 0.5--2~keV, 2--7~keV, and 0.5--7~keV.
Using our \hbox{0.5--7~keV} images, we utilized {\ttfamily wavdetect} at a
false-positive probability threshold of $10^{-5}$ to create point source
catalogs.  We converted \hbox{0.5--7~keV} point-source count-rates to fluxes
assuming an absorbed power-law spectrum with a photon index of $\Gamma = 1.5$
and Galactic extinction (see Col.~(7) in Table~1).  We treated the hot gas
component as negligible within the small area of the sources, as these galaxies
have very little diffuse emission in total, typical of low-mass ellipticals
like those studied here (e.g., O'Sullivan \etal\ 2001). Our choice of photon 
index reproduces well the mean 2--7~keV to 0.5--2~keV
count-rate ratio of the detected point sources in our sample, a value
calculated using stacking analyses (see, e.g., Lehmer \etal\ 2016 for details).

The total number of sources within each galaxy footprint that had fluxes
brighter than the $L_{\textup{X}} = 10^{38}$ erg~s$^{-1}$ limit span the range
of 1--8 (see Col.~(11)--(13) of Table~1).  Our test of the variation in the IMF
with galaxy mass is sensitive to the field LMXB population that forms within
the galactic stellar populations.  To search for potential contaminants from
unrelated \xray-detected objects, including, e.g., background active galactic nuclei (AGN), foreground
Galactic stars, central low-luminosity AGN, and LMXBs formed through dynamical
processes in GCs, we made use of archival \hst\ data (Col.(8) and (9) in Table~1).
Contaminating source populations were identified following the procedure
outlined in \S2.2 of \PeacockEtAl, which makes use of \hst\ source
colors and morphologies to identify and classify counterparts.  Regardless of
their nature, we rejected \xray\ sources with optical counterparts from further
consideration.  In total, we rejected 8 of the \xray\ sources within the
galactic footprints (see Col.~(12) and (13) of Table~1) that had optical counterparts,
which left 12 candidate field LMXBs with $L_{\rm X} > 10^{38}$~\lum\ in our sample.

%
%
\begin{figure*}
\figurenum{1}
\centerline{
\includegraphics[height=8cm]{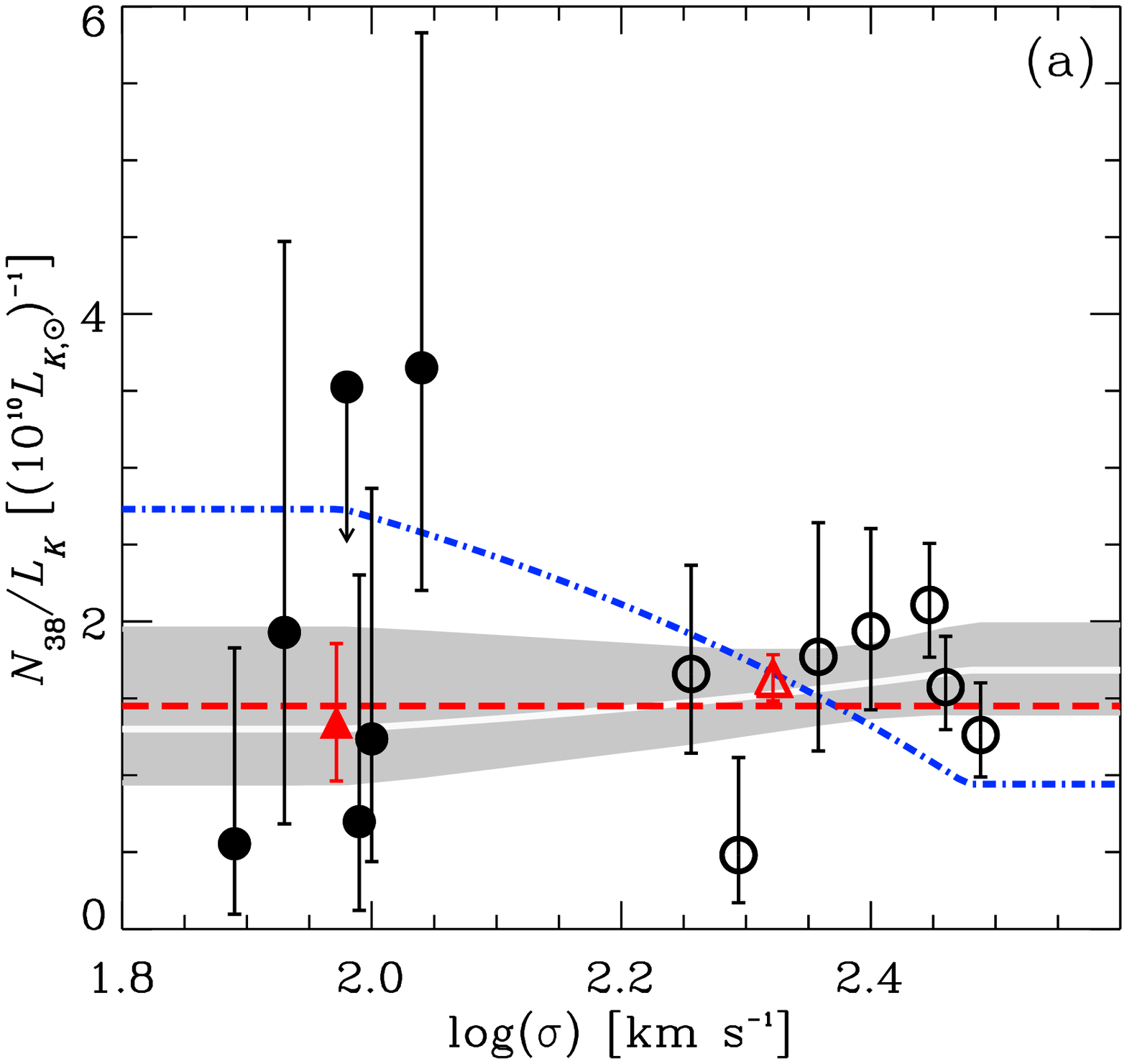} 
\hfill
\includegraphics[height=8cm]{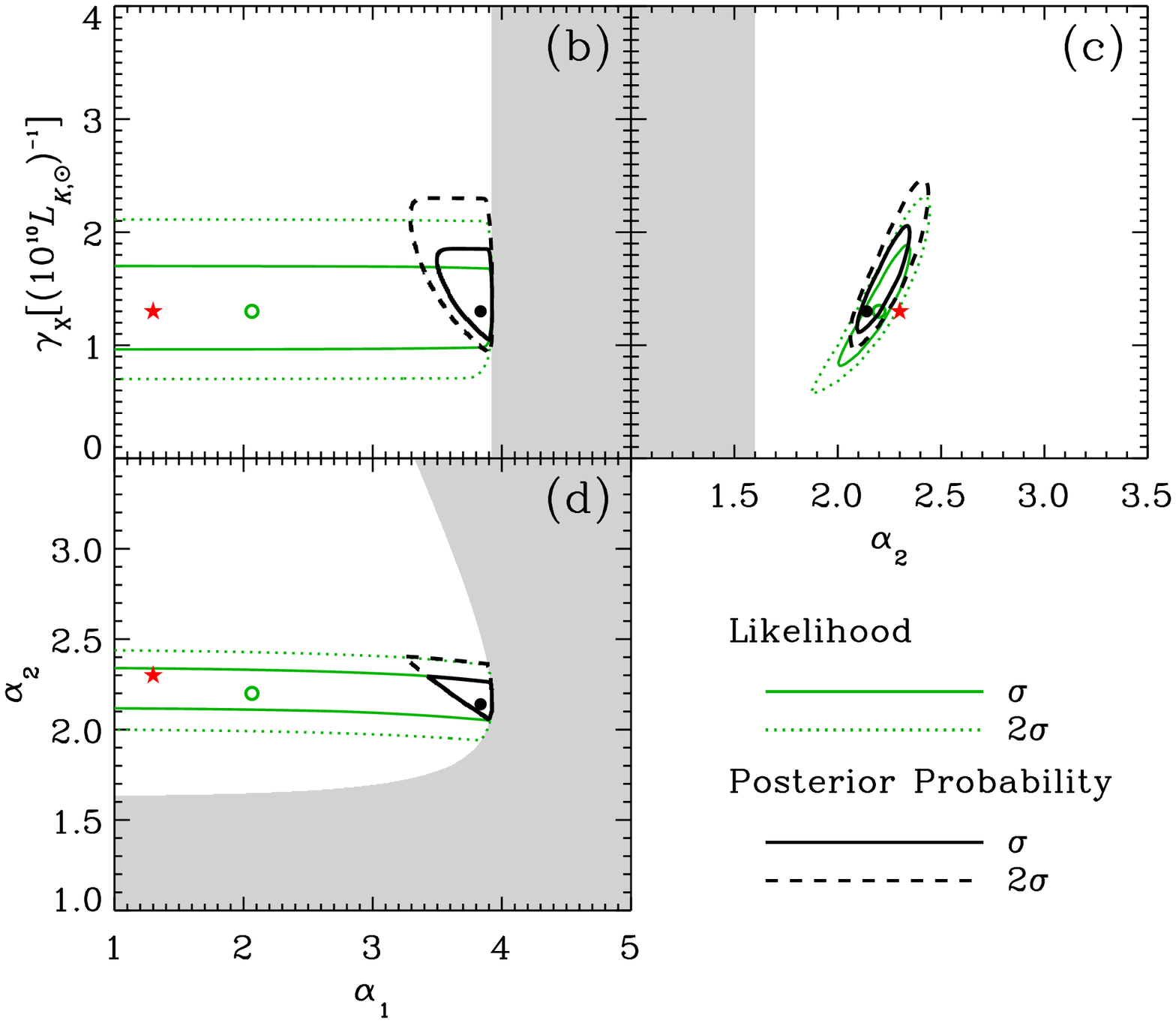} 
}
\vspace{0.1in}
\caption{
($a$)~The number of field LMXBs with $L_{\rm X} \ge 10^{38}$~\lum\ per
$10^{10}$ \lksol\ versus velocity dispersion, $\sigma$, for galaxies from this
study ({\it black filled circles}) and from \PeacockEtAl~({\it open
circles}). The red filled triangle represents the mean value from the six
galaxies in this study (see Table~1). The red open triangle represents the mean value of the
seven high-mass galaxies studied in \PeacockEtAl. Error bars are 1$\sigma$ and were
computed following Gehrels (1986). The blue dot-dash curve and red dashed line
represent, respectively, the variable and constant IMF scenarios tested by
\PeacockEtAl. The bold white curve represents our best model with our posterior
(see \S4) and interpolates between the Kroupa IMF at $\sigma
\le$~95~km~s$^{-1}$ to an IMF with $\alpha_1 =$ \bestalphaone, $\alpha_2 =$ \bestalphatwo, and
$\gamma_{\textup{X}} =$ \bestgamma\ ($10^{10}$~\lksol)$^{-1}$ at $\sigma \ge
300$~km~s$^{-1}$ (see \S3 for details). The gray shaded region is our 1$\sigma$
confidence region about our best model.
($b$)--($d$) Confidence contours for fitting parameters $\alpha_1$, $\alpha_2$,
and $\gamma_{\rm X}$ (see \S3). The green contours correspond to the likelihood
computed using only data from our LMXB study, where the black contours utilize
our LMXB data with a prior based on the study by La~Barbera \etal\ (2013; see
\S4). In each case, the solid and broken (dotted or dashed) curves represent
the 1$\sigma$ and 2$\sigma$ cuts respectively. The gray shaded region represents 
the subset of parameter space that is disallowed based on the M/L constraints 
discussed in \S3, which require each model to have an $R_{(M/L),\,i} < 3.0$. This 
results in our study discarding $\alpha_1 >= 3.9$ and $\alpha_2 <= 1.6$, and gives 
rise to the sharp cut-offs displayed in the countours. Finally, it is important to 
note that before the application of the prior, $\alpha_1$ is essentially 
unconstrained by the broad parameter space that we explored in this study. The green 
open and black filled points are our best model parameter values before and after 
the use of the prior, respectively, while the red filled stars are the reference 
Kroupa values. Values for the best fit before the prior (green 
open points) are $\alpha_1 =$ \firstalphaone, $\alpha_2 =$ \firstalphatwo, and 
$\gamma_{\textup{X}} =$ \firstgamma\ ($10^{10}$~\lksol)$^{-1}$; best fit values
after the application of the prior (black filled points) are cited above in ($a$); 
red filled stars are the Kroupa IMF parameters with $\alpha_1 = 1.3$, 
$\alpha_2 = 2.3$, and $\gamma_{\textup{X}} =$ \firstgamma\ ($10^{10}$~\lksol)$^{-1}$.}
\end{figure*}

%
\section{Results}
%
In Figure~1(a), we show the number of field LMXBs with $L_{\rm X} >
10^{38}$~\lum\ per unit $K$-band luminosity, $N_{38}/L_K$, versus velocity
dispersion, $\sigma$, for our low-mass ellipticals sample combined with the
\PeacockEtAl~sample. The ``variable'' and ``invariant'' IMF models from
\PeacockEtAl~are shown in Figure~1(a) with blue and red curves, respectively.
The variable model assumes that the IMF makes a transition from Kroupa at
$\sigma = 95$~km~s$^{-1}$ (i.e., $\log \sigma = 1.98$) to a single power-law
with slope $\alpha = 2.8$ at $\sigma = 300$~km~s$^{-1}$ (i.e., $\log \sigma =
2.48$; see details below), while the invariant model assumes a Kroupa IMF at all $\sigma$.  

At first inspection, it seems that the specific
frequency of LMXBs indicate that the IMFs of elliptical galaxies are consistent
with a single ``universal'' IMF.  However, it is important to note that the LMXB
population is tracing only the remnant population from stars with $m \simgt
8$~\msol. All that can be reliably inferred from the apparent constancy of
$N_{38}/L_K$ across $\sigma$, is that the single power law slope of the IMF for stars of $m >
0.5$~\msol~does not undergo strong variations for ellipticals of all velocity
dispersions. In fact, it may be possible that the low-mass end does vary
strongly with velocity dispersion, as found in the literature (see references
in \S1).

With this in mind, we sought to extend \PeacockEtAl~by exploring the space of 
acceptable parameters that could constrain how the IMF might vary, by constructing a suite of 
IMF models and comparing their predicted $N_{38}/L_K$ versus $\sigma$
tracks with our constraints shown in Figure~1(a).  In this process, we followed
a similar, but more generalized procedure to that outlined in \S4.2 of \PeacockEtAl. 
Below, we summarize our model.

We began by considering a broken power-law form for an IMF corresponding to
high-mass ellipticals:
%
\begin{equation}
\begin{split}
\frac{dN}{dm} = N_{0} \left \{
\begin{array}{ll}
2^{(\alpha_2 - \alpha_1)} & m^{-\alpha_{1}}\,\,\,\,\,\,\,\,\,0.1\,M_\odot < m < 0.5\,M_\odot \\
& m^{-\alpha_{2}}\,\,\,\,\,\,\,\,\,m > 0.5\,M_\odot, \\
\end{array}
  \right.
\end{split}
\end{equation}
%

\noindent where for a Kroupa IMF, $\alpha_1$ = 1.3 and $\alpha_2$ = 2.3.
$N_{0}$ is a constant of normalization, which, in our procedure, normalizes the
0.1--100~\msol\ integrated IMF to 1~\msol.  Therefore, $N_{0}$ varies with
$\alpha_1$ and $\alpha_2$.

Next, we constructed a grid of IMFs over $\alpha_1 =$~1--5 and $\alpha_2
=$~1--3.5 with 801 and 501 steps of 0.005, respectively, thus resulting in a grid of
\hbox{$n$ = 401,301} unique IMFs.  For the $i^{\rm th}$ IMF, we quantified the
$K$-band mass-to-light ratio, $(M/L_K)_i$, by running the stellar population
synthesis code {\ttfamily P\'EGASE} (Fioc \& Rocca-Volmerange~1997, 1999), adopting 
for consistency the assumption in \PeacockEtAl~of a single burst star-formation 
history of age 10~Gyr and solar metallicity. In this procedure, $M$ represents the 
total initial stellar mass and therefore does not vary with age. We defined the ratio of the $i^{\rm th}$ 
mass-to-light ratio to that of the Kroupa case as
%
\begin{equation}
R_{(M/L),\,i} = \frac{(M/L_K)_{i}}{(M/L_K)_{\textup{kro}}} = \frac{L_{K, {\rm kro}}}{L_{K, i}}.
\end{equation}
%

\noindent In order to keep $R_{(M/L)}$ consistent with observations of how the
dynamical and stellar $M/L$ varies with $\sigma$, we required that
$R_{(M/L),\,i} < 3.0$ (e.g., Cappellari \etal\ 2012; Conroy \etal\ 2012).  This
requirement on $R_{(M/L)}$ resulted in the discarding of models with specific
combinations of $\alpha_1$ and $\alpha_2$ (notably, models with $\alpha_1 \ge
3.9$ or $\alpha_2 \le 1.6$ were rejected; see Fig.~1(d)).  This constraint
limited the number of models considered to $n = 198,298$ total models, which we
utilize hereafter.

Because the prevalence of the LMXB population is sensitive to the underlying
compact object population of NSs and BHs, which are remnants of $>$~8~\msol\
stars, we calculated the number of stars per solar mass that become compact
objects for a given IMF by integrating the IMF from 8--100~\msol:
%
\begin{equation}
N_{{\rm CO}, i} = N_{0,\,i} \int_{8}^{100} m^{-\alpha_{2,i}} dm.
\end{equation}
%

\noindent where $N_{0,\,i}$ is the $i^{\rm th}$ normalization factor. These mass fractions 
allow us to compute the $K$-band luminosity normalized ratio of expected NSs and BHs generated 
by the $i^{\textup{th}}$ IMF to that of a Kroupa IMF by,
%
\begin{equation}
R_{\textup{CO},i} \equiv \frac{(N_{{\rm CO}}/L_{K})_i}{(N_{\textup{CO}}/L_{K})_{\rm kro}} =\frac{N_{{\rm CO}, i}}{N_{\textup{CO, kro}}} R_{(M/L),\,i}.
\end{equation}
%

From these quantities, we construct a variable IMF model that varies smoothly
with $\sigma$, bridging the low-mass ellipticals Kroupa IMF to the high-mass
ellipticals IMF (i.e., the $i^{\rm th}$ IMF).  We define our variable IMF function
over the range $\sigma$ = \hbox{95--300~km~s$^{-1}$}, within which we
require that the IMF varies as a function of $\sigma$ following the Cappellari
\etal\ (2013) relation of $(M/L)_\sigma \propto \sigma^{0.72}$.  Under these
assumptions, we can quantify the fraction of the variable IMF that is composed
of the $i^{\rm th}$ IMF as a function of $\sigma$ to reflect the desired result 
that galaxies with $\sigma \le$~95~km~s$^{-1}$ have no contribution from the 
$i^{\rm th}$ IMF, but will only be composed of a Kroupa IMF, while galaxies with $\sigma
\ge$~300~km~s$^{-1}$ will be solely composed of the $i^{\rm th}$ IMF component:

\begin{equation}
\begin{split}
F(\sigma) =  \left \{
  \begin{array}{ll}
 & 0 \;\;\;\;\;\;\;\;\;\;\;\;\;\;\;\;\;\;\;\;\;\;\;\;\;\; (\sigma \le 95~{\rm
km~s^{-1}})  \\[6pt]
 & \frac{\sigma^{0.72} - 95^{0.72}}{300^{0.72} - 95^{0.72}} \;\;\;\;\;\;\;\;
(95~{\rm km~s^{-1}} < \sigma < 300~{\rm km~s^{-1}})  \\[8pt]
 & 1 \;\;\;\;\;\;\;\;\;\;\;\;\;\;\;\;\;\;\;\;\;\;\;\;\;\; (\sigma \ge 300~{\rm
km~s^{-1}}),  \\
\end{array}
  \right.
\end{split}
\end{equation}
\noindent We define the complementary fraction $F_{\textup{kro}}$,
%
\begin{equation}
F_{\textup{kro}}(\sigma) = 1 - F(\sigma),
\end{equation}
%
which is then the fraction of the variable IMF that is composed of a Kroupa IMF.

We combine Equations~(4), (5), and (6) and define a composite function:
%
\begin{equation}
\begin{split}
R_{\textup{comp},i}(\sigma) & \equiv \frac{(N_{{\rm CO}}/L_{K})_{\sigma,i}}{(N_{\textup{CO}}/L_{K})_{\rm kro}} = R_{\textup{CO},i} F(\sigma) + F_{\textup{kro}}(\sigma) \\[10pt]
& = 1 - \left( 1 - R_{\textup{CO},i} \right)F(\sigma),
\end{split}
\end{equation}
%
which represents the $\sigma$-dependent number of compact objects per unit
$K$-band luminosity compared to the Kroupa IMF.  
We then arrive at a function that can be used to predict the number of observed LMXBs per unit $K$-band light:

%
\begin{equation}
\begin{split}
	\left( \frac{N_{\textup{X}}}{L_{K}} \right)_i & = \xi_i \left(\frac{N_{\textup{CO, kro}}}{L_{K, \textup{kro}}} \right)   R_{\textup{comp},i}(\sigma) \\
	& = \gamma_{{\rm X},i}  R_{\textup{comp},i}(\sigma) .
\end{split}
\end{equation}
%

Here $\xi_i$ represents the luminosity-dependent fraction of the compact object
population that is actively involved in an LMXB phase, derived using the
$i^{\rm th}$ model. $\gamma_{\textup{X},i}$ is a fitted scaling factor for the
$i^{\rm th}$ model that allows us to express this quantity as an observed
frequency of LMXBs by correcting for a portion of $N_{CO}$ which are not LMXBs 
(e.g. BH-BH pairs, BH-NS pairs, etc.). We note that we assume that $\xi$ is independent of
$\sigma$, which simplifies the computation of our IMF models, but may not be
physically accurate. A more detailed treatment requires \xray\ binary
population synthesis modeling for various IMFs (e.g., Fragos \etal\ 2013),
that would involve variations in the mass ratio distribution with IMF. However, such treatments 
are likely to introduce many parameters into the analysis for which there are no plausible physical 
models for varying binarity, and for which there are no solid empirical constraints, making the 
benefit of such modeling inconclusive. Such work is beyond the scope of this paper (see \S4).

Using Equation~(8), and the constraints on $N_{38}/L_K$ presented in Figure~1(a),
we computed maximum likelihood values for all 198,298 IMFs in our grid using the
Cash statistic ({\ttfamily cstat}; Cash~1979). This procedure resulted in a normalized likelihood cube with three dimensions:
$\alpha_1$, $\alpha_2$, and $\gamma_{\textup{X}}$. From our grid of models,
the maximum-likelihood model and 1$\sigma$ errors are \aonebefore,
\atwobefore, and \scbefore\ (see the 68\% and 95\% green contours in Fig.~1(b)--1(d)).  

As we suspected, LMXBs provide stringent constraints
on variations of the IMF for intermediate to massive stars (i.e., $m > 8$~\msol), and to the extent that we assume a single power law over the range of $m > 0.5$~\msol, a stringent constraint on $\alpha_2$. However, LMXBs
essentially provide no useful constraints on the low-mass IMF slope.

Indeed, in Figure~1(a) we can surmise that the ``invariant'' case (red curve)
is formally acceptable and within the 1 $\sigma$ threshold of our maximum
probability model, while the interpolation to a single power-law IMF with
$\alpha = 2.8$ (blue curve) is not consistent with the LMXB observations. With
an enhanced galaxy sample we not only confirm the result of \PeacockEtAl~with
better statistics, but also extend it by showing that it is possible to have an
IMF that varies from being Kroupa for low-mass ellipticals to being
bottom-heavy for high-mass ellipticals, as has been reported in the literature.
In addition, the LMXB data suggest that the single power law slope for such a varying IMF above
$0.5$~\msol~is unlikely to change much with velocity dispersion.  In the next
section, we utilize the formalism above, combined with a prior on how the IMF
mass fraction due to low-mass stars can vary with
velocity dispersion, to better constrain the values of $\alpha_1$, $\alpha_2$,
and $\gamma_{\textup{X}}$.

%
\section{Discussion and Conclusions}
%

As discussed in \S3, LMXBs provide strong constraints on the variation of the IMF for 
stars with $m > 0.5$~\msol~with velocity dispersion; however, variations in the
low-mass end of the IMF are not well constrained by LMXBs alone.  The green contours 
in Figure~1(b)--1(d) show our 1 and 2$\sigma$ confidence intervals for $\alpha_1$, $\alpha_2$, and $\gamma_{\rm X}$ 
using only LMXB data.

It is clear that these data are unable to constrain well $\alpha_1$; however, we show that
the model slope above $0.5$~\msol, $\alpha_2$, and normalization factor $\gamma_{\rm X}$, 
were strongly constrained. In order to better constrain $\alpha_1$, we utilized a prior derived by
La Barbera \etal\ (2013), who concluded that the gravity-sensitive near-IR absorption features
observed in high-mass elliptical galaxy ($\sigma \approx$~300~km~s$^{-1}$) spectra
require $\approx$70--90\% of the total initial stellar mass to be contained in stars with $m <
0.5$~\msol, as integrated directly from the IMF. From Equation~(1), we can compute the low-mass fraction for each of the $n$ IMF models defined above following
%
\begin{equation}
\begin{split}
f_i(<0.5~M_\odot) & = \frac{M(<0.5~M_\odot)_i}{M_\odot} \\
& = \frac{N_{0,i} 2^{\alpha_{2,i}-\alpha_{1,i}}}{M_\odot} \int_{0.1}^{0.5} m^{-\alpha_{1,i}+1} dm.
\end{split}
\end{equation}
%

Using the low-mass fractions calculated via Equation~(9), we next 
assigned a flat prior of 1 for $0.7 < f_i < 0.9$, and 0 elsewhere for our grid 
of IMF models. Multiplying this prior by our likelihood cube (see \S3),
and renormalizing, resulted in a posterior probability distribution which we
display as black contours in Figure~1(b)--1(d).  The resulting posterior probability
provides a stringent constraint on $\alpha_1$ for massive ellipticals, resulting in best model
values of \aoneafter, \atwoafter, and \scafter. 
The white curve, and gray 1$\sigma$ envelope, displayed in Figure~1(a) shows
our best model, which smoothly varies between a Kroupa IMF for low-mass 
ellipticals, to a broken power-law IMF for high-mass ellipiticals. The high-mass 
galaxy IMF component has a steep slope of $\alpha_1 =$ \bestalphaone\ for stars $< 0.5$~\msol, and a slope of 
$\alpha_2 =$ \bestalphatwo\ for stars $\ge 0.5$~\msol. We note that while $\alpha_2$ is slightly flatter than
the slope given by Kroupa, it still squarely falls 
within the uncertainty of the Kroupa IMF ($\alpha_2 = 2.3 \pm 0.7$; Kroupa, 2001).

By construction, this result is consistent with both the IMF for massive
ellipticals being bottom-heavy, as inferred in the literature (see references
in \S1), and the IMF not varying significantly across the mass spectrum at the
high-mass end, as inferred by the LMXB populations (see also \PeacockEtAl).
Additionally, the variable IMF has been constructed to yield a total (including
stellar remnants) mass-to-light ratio that varies with velocity dispersion
following the observational constraints from Cappellari \etal\ 2013.  In
absolute terms, the stellar mass-to-light ratio of our best-fit model, varies
from $M_\star/L_K$ = \mstarlow\ at $\sigma = 90$~km~s$^{-1}$ to
$M_\star/L_K$ = \mstarhigh\ at $\sigma = 300$~km~s$^{-1}$. The initial mass-to-light ratios, 
(initial stellar mass over $K$-band luminosity) are $M/L_K$ = \MLlow\ at 
$\sigma = 90$~km~s$^{-1}$ to $M/L_K$ = \MLhigh\ at $\sigma = 300$~km~s$^{-1}$, 
resulting in a ratio $R_{M/L} = 2.78$. These values are below the limits placed on 
the dynamical mass-to-light ratio values measured in the literature (e.g., Cappellari 
et al. 2013; Conroy et al. 2013).

We also noted in \S1 that recent studies have found evidence for radial gradients in 
the IMFs of massive ellipticals, in which a bottom-heavy IMF appears to be appropriate for 
the inner regions of the galaxy, while a more Kroupa-like IMF is appropriate in the outer 
regions (see, e.g., Mart\'in-Navarro et al. 2015; La Barbera et al. 2016a, 2016b). We tested 
to see whether the LMXB population showed evidence for such radial gradients by calculating 
the average $N_{38}/L_{K}$ values for LMXBs that were located within and outside of the 
effective radii, $r_e$, of the massive elliptical galaxy population. We utilized $r_e$ values 
from Cappellari et al. (2013) to divide the $L_{X} > 10^{38}$~erg~s$^{-1}$ LMXB catalogs from 
Peacock et al.~(2014) into inner- and outer-LMXB populations, and used the 2MASS $K$-band 
images to calculate the fraction of $L_K$ that was within and outside of $r_e$. We found 
consistent values of $N_{38}/L_{K} = 1.4 \pm 0.5$ and $1.7 \pm 0.6$ for the regions within 
and outside $r_e$, respectively, indicating that such gradients do not have a significant effect 
on our results.

To emphasize, our likelihoods are broadly insensitive to $\alpha_1$. We relied on 
La Barbera et al. (2013) as a meaningful prior and showed we are able to reconcile 
their results with our analysis, with a statistically insignificant change in the 
derived values for the other two parameters, $\alpha_2$ and $\gamma_{\rm X}$.
A choice of a different, meaningful prior would necessarily entail a different posterior 
probability density for $\alpha_1$, however, the essential constraint that this 
study places on $\alpha_2$ would not change significantly.

If a slight flattening of the high-mass slope of the 
generalized IMF is common in high-mass elliptical galaxies (shown to be required if we 
accept that the bottom-heavy IMF claims in the literature are correct), then 
several additional predictions result. We would expect a factor of $\sim$2 
higher supernova rate in the high redshift versions of these galaxies than their 
present day stellar masses imply. A more severe effect might be seen on the rate 
of long duration gamma-ray bursts (GRBs), which are often thought to come from only 
the most massive stellar explosions. Presently, we have limited information 
about very high-redshift GRBs, and almost no information about very high-redshift 
supernovae (SNe). The information that exists suggests a GRB rate not easily explained by 
detectable star forming galaxies (Tanvir et al. 2012), but the present results are easily 
explained by having most of the star formation take place in relatively small galaxies, 
below the \hst~detection threshold. We would expect similar increases in the rates of double 
neutron star mergers producing short GRBs and gravitational wave sources, 
as well as an enhancement, albeit a somewhat lower one, of the rate of Type Ia 
SNe, since more white dwarfs would be produced and they would be skewed more heavily 
toward the massive end of the white dwarf spectrum.

It is also important to note that if the slope of the high-mass end of the IMF changes by 
about 0.1 dex, then the ratio of core collapse supernovae from $\approx$40~\msol\ stars to that 
from $\approx$8~\msol\ stars changes by a factor of about 20\%. The yields from the most massive 
core collapse supernovae to those from the least massive ones can vary quite dramatically 
(e.g. Kobayashi \etal\ 2006). As a result, interstellar medium abundances may provide a 
complementary test of the IMF. Because additional metal enrichment comes from thermonuclear 
supernovae, classical novae, and mass loss, a detailed treatment of this problem must 
be undertaken before a clear prediction can be made.

Future \xray\ and high-resolution optical observations, and new population
synthesis analyses could substantially improve constraints on the IMF
variation with velocity dispersion.  In particular, new \chandra\ and \hst\
observations of low--to--intermediate mass ellipticals would be helpful in
ruling-out a scenario where the high-mass end of the IMF is constant across all
$\sigma$.  Deeper \chandra\ observations of the low-mass galaxies in this study
would have a similar effect, allowing us to probe to the more numerous
population of low-luminosity LMXBs.  In a forthcoming paper, Peacock \etal\
(in-prep), will be examining the $L_{\rm X} \simgt 10^{37}$~\lum\ field LMXB
population in low-mass elliptical NGC~7457 and comparing it with similar
binaries detected in deep observations of massive ellipticals.  

Most effectively testing the variations of the IMF in elliptical galaxies will
require simultaneous modeling of near-IR spectroscopic data along with
LMXB constraints using the combination of stellar population synthesis and
\xray\ binary population synthesis modeling (e.g., Fragos \etal\ 2013; Madau \&
Fragos~2016), in which the IMF, stellar ages, metallicities, and other physical
parameters that influence \xray\ binary formation are all modeled self-consistently. 
Such a population synthesis framework will be an important
future step for advancing our knowledge of how the IMF varies among elliptical galaxies.

Additionally, a further understanding of the details of the stellar populations 
in the near-IR must be developed to ensure that the claims of a 
bottom-heavy IMF are reliable. At the present time, stellar evolution codes used 
for these purposes have limited or no treatment of unusual classes of stars such as 
interacting binaries and the products of binary interactions. Such stars are likely 
to be relatively unimportant for tracing out the optical bands where the models have 
been best calibrated, however, given that the largest radius stars are most likely to interact, 
these stars may be increasingly important toward the reddest parts of the spectral 
energy distribution where red giants and asymptotic giant branch stars are most 
important. For instance, it has already been shown that the S-type stars might appear 
with different frequencies in larger and smaller elliptical galaxies, and can potentially 
mimic the effects of bottom-heavy IMFs on the Wing-Ford band and the Na~I D line, although 
not on Ca~II (Maccarone 2014). 

Given the coincidence needed between changes in both the high- and low-mass ends of the IMF 
to reproduce the results we see here, as well as the subtlety of the features that have been used 
to suggest the bottom-heavy IMF, more work to investigate further the level of systematic 
effects from stars not included in standard stellar population synthesis models would be 
well-motivated.

%
\acknowledgements
%

D.A.C., B.D.L., and R.T.E. gratefully
acknowledge support from \chandra\ \xray\ Center (CXC) grant GO4-15090A.  D.A.C. acknowledges support from USRA and GSFC, as well as generous support from the University of Arkansas. A.K. acknowledges CXC grants GO4-15090C and GO5-16084B, and T.M. awknowledges GO4-15090B. M.P. and S.Z. acknowledge support from NASA ADAP grant NNX15AI71G, and M.P. additionally acknowledges CXC grant GO5-16084A and the {\itshape Hubble Space Telescope} (\hst\/) grant HST-GO-13942.001-A.

%

%

\end{document}